\documentclass{article}
\usepackage{spconf,amsmath,graphicx}
\usepackage{amsthm,amsmath,amssymb}
\usepackage{mathrsfs}
\usepackage{booktabs}
\usepackage{threeparttable}
\usepackage{multirow,multicol}
\usepackage{makecell}

\title{Transformer-based Online CTC/attention End-to-End Speech Recognition Architecture}
\name{Haoran Miao$^{1,2}$, Gaofeng Cheng$^{1}$, Changfeng Gao$^{1,2}$, Pengyuan Zhang$^{1,2}$, Yonghong Yan$^{1,2,3}$\thanks{This work is partially supported by the National Key Research and Development Program (Nos. 2018YFC0823402, 2018YFC0823401, 2018YFC0823405, 2018YFC0823400), the National Natural Science Foundation of China (Nos. 11590774, 11590772, 11590770),the Key Science and Technology Project of the Xinjiang Uygur Autonomous Region (No.2016A03007-1).}}
\address{$^1$Key Laboratory of Speech Acoustics and Content Understanding, Institute of Acoustics, China\\  $^2$University of Chinese Academy of Sciences, China\\  $^3$Xinjiang Laboratory of Minority Speech and Language Information Processing, Xinjiang Technical\\ Institute of Physics and Chemistry, Chinese Academy of Sciences, China}
\begin{document}
\ninept
\maketitle
\begin{abstract}
Recently, Transformer  has gained success in automatic speech recognition (ASR) field. However, it is challenging to deploy a Transformer-based end-to-end (E2E) model for online speech recognition.
In this paper, we propose the Transformer-based online CTC/attention E2E ASR architecture, which contains the chunk self-attention encoder (chunk-SAE) and the monotonic truncated attention (MTA) based self-attention decoder (SAD). 
Firstly, the chunk-SAE splits the speech into isolated chunks. To reduce the computational cost and improve the performance, we propose the state reuse chunk-SAE.
Sencondly, the MTA based SAD truncates the speech features monotonically and performs attention on the truncated features.
To support the online recognition, we integrate the state reuse chunk-SAE and the MTA based SAD into online CTC/attention architecture. 
We evaluate the proposed online models on the
HKUST Mandarin ASR benchmark and achieve a $23.66\%$ character error rate (CER) with a 320 ms latency. Our online model yields as little as $0.19\%$ absolute CER degradation compared with the offline baseline, and achieves significant improvement over our prior work on Long Short-Term Memory (LSTM) based online E2E models.\par
\end{abstract}
\section{Introduction}
\label{sec:intro}

In recent years, the end-to-end (E2E) automatic speech recognition (ASR) has gained popularity in ASR community \cite{CTC_Graves_2006,attention_bengio_2015,LAS_oriol_2016,amodei2016deep,seq2seq_state_of_the_art_2018_google,watanabe2017hybrid}. E2E ASR models simplify the hybrid DNN/HMM ASR models by replacing the acoustic, pronunciation and language models with one single deep neural network, and thus transcribe speech to text directly.
To date, E2E ASR models have achieved significant improvement in ASR field \cite{amodei2016deep, seq2seq_state_of_the_art_2018_google, watanabe2017hybrid}. The hybrid Connectionist Temporal Classification (CTC) / attention E2E ASR architecture \cite{watanabe2017hybrid} has attracted lots of attention because it combines the advantages of CTC models and attention models. During training, the CTC objective is attached to the attention-based encoder-decoder model as an auxiliary task. During decoding, the joint CTC/attention decoding approach is adopted in the beam search \cite{CTC-Attention-ACL-2017}.
However, it is difficult to deploy the online CTC/attention E2E ASR architecture because of global attention mechanisms \cite{Bahdanau2014Neural} and CTC prefix scores \cite{watanabe2017hybrid, kawakami2008supervised}, which depend on the entire input speech.
Our prior work \cite{Miao2019,mta} has streamed this architecture from both the model structure and decoding algorithm aspects.
On the model structure aspect, we proposed the stable monotonic chunk-wise attention (sMoChA) \cite{Miao2019} and monotonic truncated attention (MTA) \cite{mta} to stream attention mechanisms, and applied the latency-controlled bidirectional long short-term memory (LC-BLSTM) as the low-latency encoder.
On the decoding aspect, we proposed the online joint decoding approach, which includes truncated CTC (T-CTC) prefix scores and dynamic waiting joint decoding (DWJD) algorithm \cite{Miao2019}. \par
Recently, Transformer \cite{Vaswani2017} has gained success in ASR field \cite{Dong2018speechtransformer,Karita2019,Pham2019}.
Transformer-based models are parallelizable and competitive to recurrent neural networks \cite{Karita2019A}. However, the vanilla Transformer is inapplicable to online tasks for two reasons: First, the self-attention encoder (SAE) computes the attention weights on the whole input frames; Second, the self-attention decoder (SAD) computes the attention weights on the whole outputs of SAE.

In this paper, we stream the Transformer and integrate it into the CTC/attention E2E ASR architecture.
On the SAE aspect, we propose the chunk-SAE which splits the input speech into isolated chunks of fixed length. Inspired by Transformer-XL \cite{Dai2019}, we further propose the state reuse chunk-SAE which reuses the stored states of the previous chunks to reduce the computational cost.
On the SAD aspect, we propose the MTA based SAD, which performs attention on the truncated historical outputs of SAE.
Finally, we propose the Transformer-based online CTC/attention E2E ASR architecture via the online joint decoding approach \cite{Miao2019}. Our experiments shows that the proposed online model with a 320 ms latency achieves 23.66\% character error rate (CER) on HKUST, with only 0.19\% absolute CER degradation compared with the offline baseline.

The rest of this paper is organized as follows. In Section~\ref{sec:online}, we describe the online CTC/attention E2E architecture proposed in our prior work \cite{Miao2019, mta}. In Section~\ref{sec:trans}, we introduce the Transformer architecture. In Section~\ref{sec:stream}, we describe the online Transformer-based CTC/attention architecture. The experiments and conclusions are presented in Sections~\ref{sec:exp} and \ref{sec:con}, respectively.\par

\section{Online CTC/attention E2E Architecture}
\label{sec:online}

In our prior work \cite{Miao2019}, we proposed an online hybrid CTC/attention E2E ASR architecture, which consists of the LC-BLSTM encoder, sMoChA and LSTM decoder. During training, we introduce the CTC objective as an auxiliary task, and the loss function is defined by:
\begin{equation}
	\mathcal{L} = \alpha\mathcal{L}_{\rm dec} + (1-\alpha)\mathcal{L}_{\rm ctc},\label{train}
\end{equation}
where $\alpha$ is a hyperparameter, $\mathcal{L}_{\rm dec}$ and $\mathcal{L}_{\rm ctc}$ are loss functions from the decoder and CTC.
During decoding, we adopt the online joint decoding approach, which is defined by:
\begin{eqnarray}
	\hat{Y}={\rm arg}\max_{Y\in\mathcal{Y}^*}\{\lambda\log P_{\rm dec}(Y|X)\!&+&\!(1-\lambda)\log P_{\rm t\textit{-}ctc}(Y|X) \notag\\
															&+& \gamma\log P_{\rm lm}(Y)\}, \label{decode}
\end{eqnarray}
where $P_{\rm dec}(Y|X)$ and $P_{\rm t\textit{-}ctc}(Y|X)$ are the probabilities of the hypothesis $Y$ conditioned on input frames $X$ from the decoder and T-CTC \cite{Miao2019}, and $P_{\rm lm}(Y)$ is the language model probability. The hyperparameters $\lambda$ and $\gamma$ are tunable. For online decoding, we proposed DWJD algorithm \cite{Miao2019} to 1) coordinate the forward propagation in the encoder and the beam search in the decoder; 2) address the unsynchronized predictions of the sMoChA-based decoder and CTC outputs. \par

MTA \cite{mta}, which performs attention on top of the truncated historical encoder outputs, outperforms the sMoChA by exploiting longer history. Formally, we denote ${\rm \mathbf{q}}_{i}$ and ${\rm \mathbf{h}}_{j}$ as the $i$-th decoder state and the $j$-th encoder output, respectively. Similar to monotonic chunk-wise attention \cite{Chiu2018mocha}, MTA defines the probability $p_{i,j}$ of truncating encoder outputs at ${\rm \mathbf{h}}_{j}$ as:
\begin{equation}
    p_{i,j} = {\rm Sigmoid}(g\frac{{\rm \mathbf{v}}^{\top}}{||{\rm \mathbf{v}}||}{\rm tanh}({\rm \mathbf{W}}_1{\rm \mathbf{q}}_{i-1}+{\rm \mathbf{W}}_2{\rm \mathbf{h}}_j+{\rm \mathbf{b}})+r), \label{MTA_p}
\end{equation}
where the matrices ${\rm \mathbf{W}}_1$, ${\rm \mathbf{W}}_2$, vectors ${\rm \mathbf{b}}$, ${\rm \mathbf{v}}$ and scalars $g$, $r$ are trainable parameters. Then, the attention weight $a_{i,j}$ is computed by:
\begin{equation}
    a_{i,j} = p_{i,j}\prod\nolimits^{j-1}_{k=1}(1-p_{i,k}), \label{MTA_a}
\end{equation}
where $a_{i,j}$ indicates the probability of truncating encoder outputs at ${\rm \mathbf{h}}_{j}$ and skipping the encoder outputs before ${\rm \mathbf{h}}_{j}$. During decoding, MTA determines a truncation end-point $t_{i}$ for the $i$-th decoder step by:
\begin{equation}
    z_{i,j} = \mathbb{I}(p_{i,j}>0.5\land j\geq t_{i-1}), \label{MTA_z}
\end{equation}
where $z_{i,j}$ denotes the indicator of truncating or do not truncating encoder outputs at ${\rm \mathbf{h}}_{j}$, and $\mathbb{I}$ represents an indicator function. By the condition $j\geq t_{i-1}$ in Eq.~\ref{MTA_z}, MTA enforces the end-point to move in a left-to-right mode. Once $z_{i,j}=1$ for some $j$, MTA sets $t_{i}$ to $j$. Finally, MTA performs attention on the truncated encoder outputs:
\begin{equation}
    {\rm \mathbf{r}}_{i} = \sum\nolimits^{t_{i}}_{j=1}a_{i,j}{\rm \mathbf{h}}_j, \label{MTA_r}
\end{equation}
where ${\rm \mathbf{r}}_{i}$ is the letter-wise hidden vector for the $i$-th decoder step.\par
During training, MTA performs attention on the whole encoder outputs:
\begin{equation}
    {\rm \mathbf{r}}_{i} = \sum\nolimits^{T}_{j=1}a_{i,j}{\rm \mathbf{h}}_j, \label{MTA_r}
\end{equation}
where $T$ denotes the number of encoder outputs. \par

\section{Transformer Architecture}
\label{sec:trans}

Transformer \cite{Vaswani2017} follows the encoder-decoder architecture using stacked self-attention and position-wise feed-forward layers for both the encoder and decoder. We briefly introduce the Transformer architecture in this section. \par

\subsection{Multi-head attention}

Transformer adopts the scaled dot-product attention to map a query and a set of key-value pairs to an output as:
\begin{equation}
    {\rm Attention}({\rm \mathbf{Q}}, {\rm \mathbf{K}}, {\rm \mathbf{V}}) = {\rm Softmax}(\frac{{\rm \mathbf{Q}}{\rm \mathbf{K}}^{\top}}{\sqrt{d_m}}){\rm \mathbf{V}},\label{scaled-dot-product}
\end{equation}
where the matrices ${\rm \mathbf{Q}}\in\mathbb{R}^{n\times d_m}$, ${\rm \mathbf{K}}\in\mathbb{R}^{m\times d_m}$ and ${\rm \mathbf{V}}\in\mathbb{R}^{m\times d_m}$ denote queries, keys and values, $n$ and $m$ denote the number of queries and keys (or values), and $d_m$ denotes representation dimension. \par
Instead of performing a single attention function, Transformer uses multi-head attention that jointly learns diverse relationships between queries and keys from different representation sub-spaces as follows:
\begin{equation}
    {\rm MultiHead}({\rm \mathbf{Q}}, {\rm \mathbf{K}}, {\rm \mathbf{V}}) = {\rm Concat}({\rm head}_1, \cdot\cdot\cdot, {\rm head}_H){\rm \mathbf{W}}^O, \label{MultiHead}
\end{equation}
\begin{equation}
    {\rm head}_h = {\rm Attention}({\rm \mathbf{Q}}{\rm \mathbf{W}}^Q_h, {\rm \mathbf{K}}{\rm \mathbf{W}}^K_h, {\rm \mathbf{V}}{\rm \mathbf{W}}^V_h),    
\end{equation}
where $H$ denotes the head number and $d_k=d_m/H$. The matrices ${\rm \mathbf{W}}^O\in\mathbb{R}^{d_m\times d_m}$ and ${\rm \mathbf{W}}^{Q,K,V}_h\in\mathbb{R}^{d_m\times d_k}$ are trainable parameters. \par


Because Transformer lacks of modeling the sequence order, the work in \cite{Vaswani2017} suggested to use sine and cosine functions of different frequencies to perform the positional encoding. \par

\subsection{Self-attention encoder (SAE)}
\label{ssec:sae}

The SAE consists of a stack of identical layers, each of which has two sub-layers, i.e. one self-attention layer and one position-wise feed-forward layer. The inputs of the SAE are acoustic frames in ASR tasks. The self-attention layer employs multi-head attention, in which the queries, keys and values are inputs of the previous layer. Besides, the SAE uses residual connections \cite{7780459} and layer normalization \cite{BaKH16} after each sub-layer.\par

\subsection{Self-attention decoder (SAD)}
\label{ssec:sad}
The SAD also consists of a stack of identical layers, each of which has three sub-layers, i.e. one self-attention layer, one encoder-decoder attention layer and one position-wise feed-forward layer. The inputs of the SAD are embeddings of right-shifted output labels. To prevent the access to the future output labels in the self-attention, the subsequent positions are masked. In the encoder-decoder attention, the queries are current layer inputs while the keys and values are SAE outputs. Besides, the SAD also uses residual connections and layer normalization after each sub-layer.\par

\section{Transformer-based Online CTC/attention E2E Architecture}
\label{sec:stream}
\begin{figure}[t]
    \centering
    \includegraphics[width=0.5\linewidth]{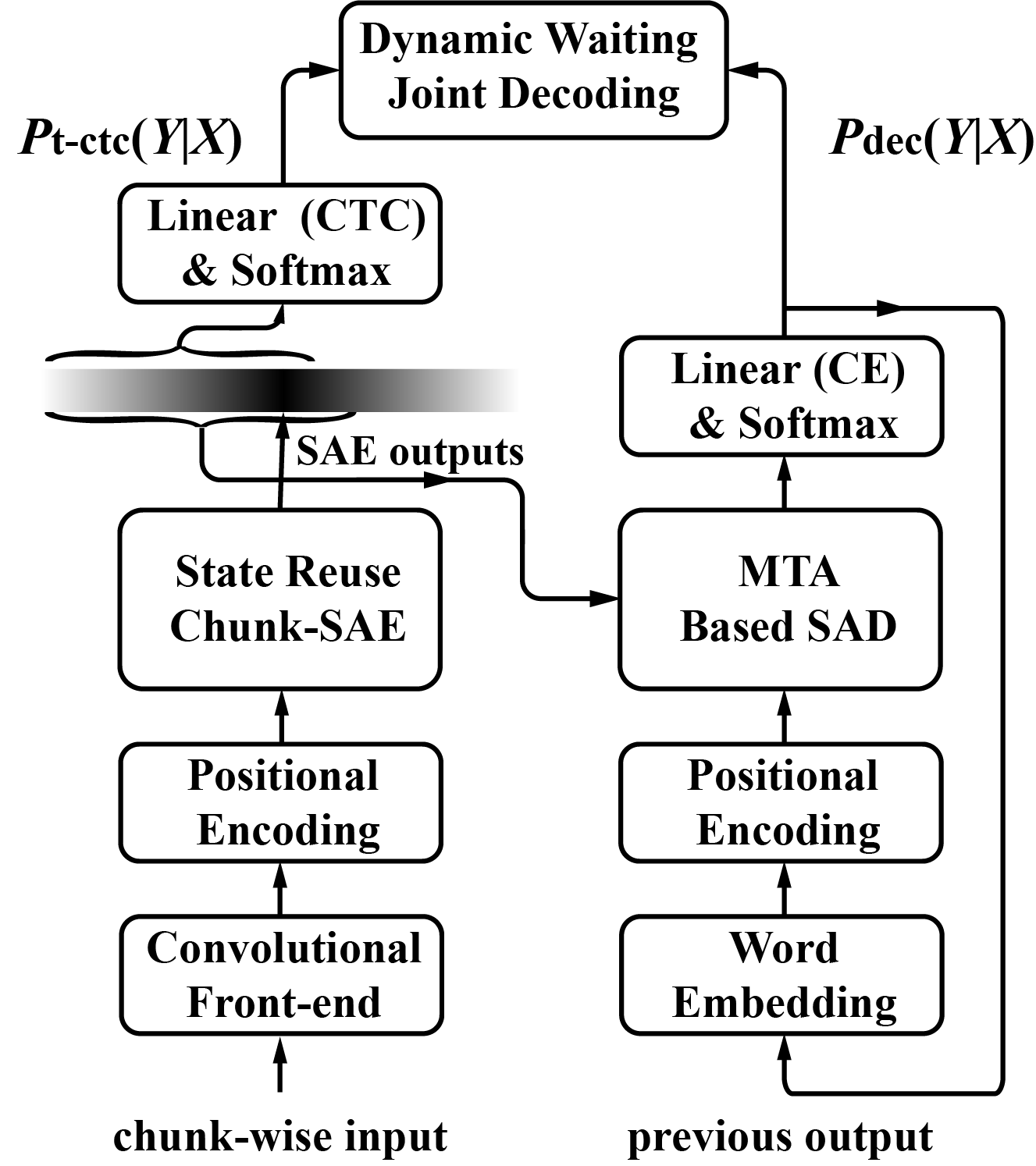}
    \caption{Transformer-based online CTC/attention E2E architecture.}
    \label{fig:model}
\end{figure}
In this section, we propose the Transformer-based online E2E model, which consists of the chunk-SAE with or without reusing stored states and MTA based SAD. The Transformer-based online CTC/attention E2E architecture is shown in Fig.~\ref{fig:model}. \par

\subsection{Chunk-SAE}

To stream the SAE, we first propose the chunk-SAE, which splits a speech into non-overlapping isolated chunks of $N_c$ central length. To acquire the contextual information, we splice $N_l$ left frames before each chunk as historical context and $N_r$ right frames after it as future context. The spliced frames only act as contexts and give no output.
With the predefined parameters $N_c$, $N_l$ and $N_r$, the receptive field of each chunk-SAE output is restricted to $N_l+N_c+N_r$ and the latency of the chunk-SAE is limited to $N_r$. 

\subsection{State reuse chunk-SAE}
\label{ssec:c-sae}
\begin{figure}[t]
    \centering
    \includegraphics[width=0.9\linewidth]{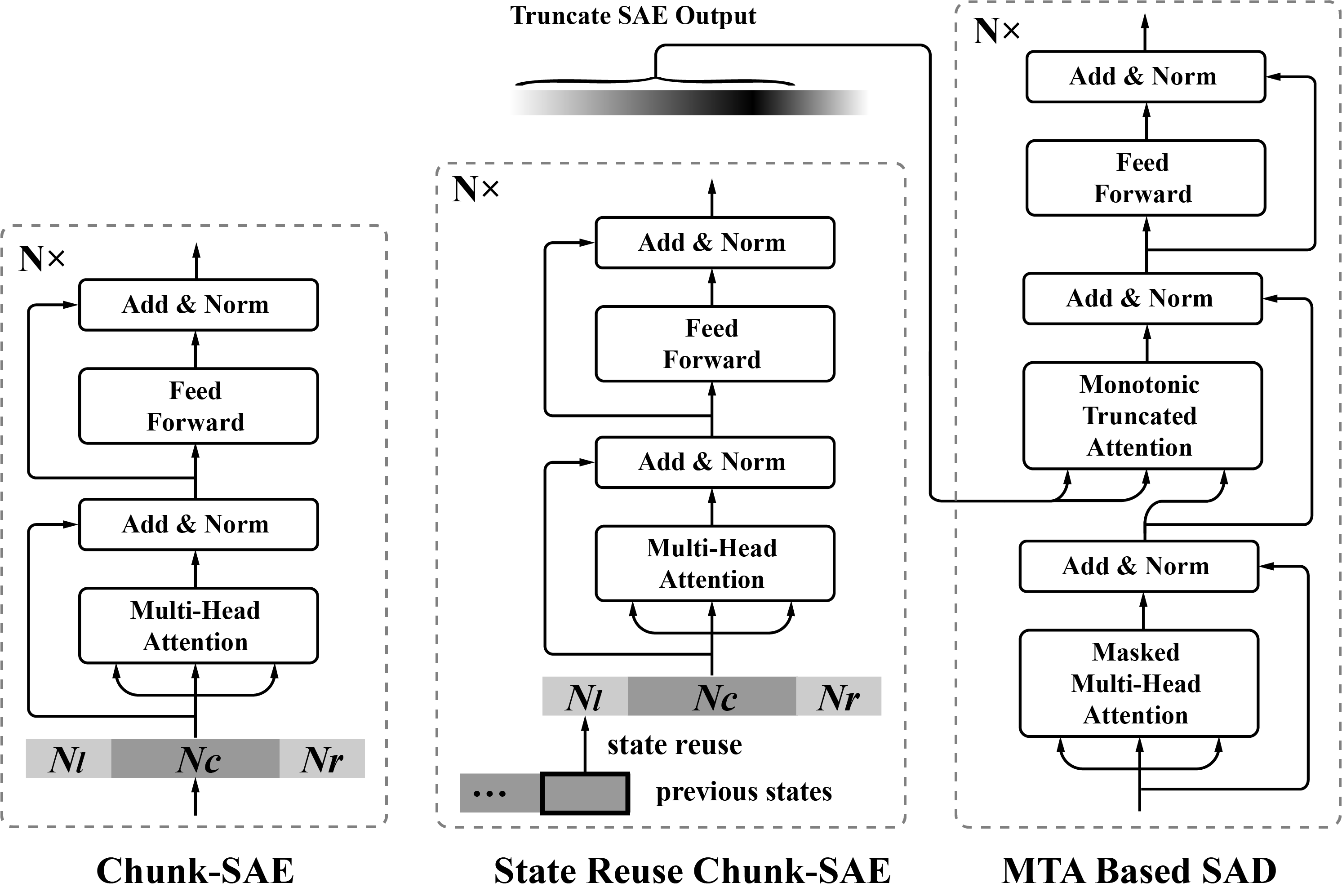}
    \caption{Illustrations of the chunk-SAE, state reuse chunk-SAE and MTA based SAD.}
    \label{fig:sae_sad}
\end{figure}

In the chunk-SAE, the historical context is re-computed for each chunk. To reduce the computational cost, we store the computed hidden states in central context. Then, when computing the new chunk, we reuse stored hidden states from the previous chunks at the same positions as historical context, which is inspired by Transformer-XL \cite{Dai2019}. Fig.~\ref{fig:sae_sad} illustrates the difference between the chunk-SAE with or without reusing hidden states. Formally, ${\rm \mathbf{s}}_{\tau}^{l}\in\mathbb{R}^{N_l\times d_m}$ and ${\rm \mathbf{h}}_{\tau}^{l}\in\mathbb{R}^{(N_c\!+\!N_r)\times d_m}$ denote the stored and newly-computed hidden states for the $\tau$-th chunk in the $l$-th layer, respectively. Then, the queries, keys and values for the $\tau$-th chunk in the $l$-th self-attention layer are defined as follows:
\begin{eqnarray}
    {\rm \mathbf{Q}}_{\tau}^{l}, {\rm \mathbf{K}}_{\tau}^{l}, {\rm \mathbf{V}}_{\tau}^{l} &=& {\rm \mathbf{h}}_{\tau}^{l-1}, {\rm \mathbf{\widetilde{h}}}_{\tau}^{l-1}, {\rm \mathbf{\widetilde{h}}}_{\tau}^{l-1}, \label{SR-CTE}\\
    {\rm where\quad \mathbf{\widetilde{h}}}_{\tau}^{l-1} &=& {\rm Concat}({\rm SG}({\rm \mathbf{s}}_{\tau}^{l-1}),\ {\rm \mathbf{h}}_{\tau}^{l-1}). \label{SR}
\end{eqnarray}
In Eq.~\ref{SR}, the function ${\rm SG}(\cdot)$ stands for stop-gradient. Therefore, the complexity of the state reuse chunk-SAE is reduced by a factor of $N_l/(N_l+N_c+N_r)$. \par
Moreover, the state reuse chunk-SAE captures long-term dependency beyond the chunks. Suppose the state reuse chunk-SAE consists of $L$ layers, the receptive field on the left side extends to as far as $L\cdot N_{l}$ frames, which is much broader than that of chunk-SAE.\par

\subsection{MTA based SAD}

To stream the SAD, we propose the MTA based SAD to truncate the receptive field in a monotonic left-to-right way and perform attention on the truncated outputs of SAE. Specifically, we substitute MTA for the encoder-decoder attention in each SAD layer, as shown in Fig.~\ref{fig:sae_sad}. Suppose the representation dimension is $d_m$, MTA performs in parallel during training as follows: \begin{equation}
    {\rm MTA}({\rm \mathbf{Q}}, {\rm \mathbf{K}}, {\rm \mathbf{V}}) = (\mathbf{P}\odot {\rm cumprod}(\mathbf{1}-\mathbf{P})){\rm \mathbf{V}}{\rm \mathbf{W}}_v,\label{MTA-TD} 
\end{equation}
\begin{equation}
    \mathbf{P} = {\rm sigmoid}(\frac{{\rm \mathbf{Q}}{\rm \mathbf{W}}_q{{\rm \mathbf{W}}_k}^{\top}{{\rm \mathbf{K}}}^{\top}}{\sqrt{d_m}}+r+\varepsilon),\label{MTA-TD-P}
\end{equation}
where the matrices ${\rm \mathbf{W}}_{\cdot}\in\mathbb{R}^{d_m\times d_m}$ and scalar bias $r$ are trainable parameters, and $\epsilon$ denotes the noise.
We define $\mathbf{P}\!=\!\{p_{i,j}\}$ as the truncation probability matrix, where $p_{i,j}$ indicates the probability of truncating the $j$-th SAE output in order to predict the $i$-th output label. In Eq.~\ref{MTA-TD}, the cumulative product function ${\rm cumprod(\mathbf{x})}=[1, x_1, x_{1}x_{2}, \cdot\cdot\cdot, \prod_{k=1}^{|x|-1}x_{k}]$ and ${\rm cumprod(\cdot)}$ applies to the rows of $\mathbf{P}$. The notation $\odot$ indicates the element-wise product. \par
MTA learns the appropriate offset for the pre-sigmoid activations in Eq.~\ref{MTA-TD-P} via the trainable scalar $r$. To prevent ${\rm cumprod}(\mathbf{1}-\mathbf{P})$ from vanishing to zeros, we initialize $r$ to a negative value, e.g. $r=-4$ in our experiments. To encourage the discreteness of the truncation probabilities, we simply add zero-mean, unit-variance Gaussian noise $\varepsilon$ to the pre-sigmoid activations only during training. \par
During decoding, we have to compute the elements in $\mathbf{P}^l\!=\!\{p_{i,j}^l\}$ row by row, where $\mathbf{P}^l$ is the truncation probability matrix in the $l$-th layer. we define $t_i^l$ as the truncation end-point belonging to the $l$-th layer when predicting the $i$-th output label. Then, the end-point is determined by:
\begin{equation}
    z_{i,j}^l = \mathbb{I}(p_{i,j}^l>0.5\land j\geq t_{i-1}^l), \label{MTA_SAD_z}
\end{equation}
where $z_{i,j}^l$ denotes the indicator of truncating or do not truncating $j$-th SAE output in $l$-th layer and $\mathbb{I}$ represents an indicator function. Once $z_{i,j}^l=1$ for some $j$, we set $t_{i}^l$ to $j$, which means that the receptive field of the $l$-th layer is restricted to $t_{i}^l$ SAE outputs.
Suppose the MTA based SAD consists of $L$ layers, there will be $L$ end-points at each decoding step. The number of truncated SAE outputs in each layer will not affect other layers. Therefore, we define the the maximum of $L$ end-points as the receptive field of the MTA based SAD. \par

\section{Experiments}
\label{sec:exp}

\subsection{Corpus}
\label{ssec:corp}

We evaluated our models using HKUST Mandarin Chinese conversational telephone \cite{Liu2006HKUST}. The HKUST consists of about 200 hours \emph{train} set for training and about 5 hours test set. We extracted 4000 utterances from the \emph{train} set as our development set. To improve the recognition accuracy, we applied the speed perturbation on the rest \emph{train} set by factors 0.9 and 1.1.\par

\subsection{Model descriptions}
\label{ssec:model}

We built all the online models using ESPnet toolkit \cite{Watanabe2018ESPnet}. For the input, we used 83-dimensional features, including 80-dimensional filter banks, pitch, delta-pitch and Normalized Cross-Correlation Functions. The features were computed with a 25 ms window and shifted every 10 ms. For the output, we adopted a 3655-sized vocabulary set, including 3623 Chinese Mandarin characters, 26 English characters, as well as 6 non-language symbols denoting laughter, noise, vocalized noise, blank, unknown-character and sos/eos. \par
We used 2-layer convolutional neural networks (CNN) as the front-end. Each CNN layer had 256 filters, each of which has $3\times 3$ kernel size with $2\times 2$ stride, and thus the time reduction of the front-end was $1/4$. The SAE and SAD had 12 and 6 layers, respectively. All sub-layers, as well as embedding layers, produced outputs of dimension 256. In the multi-head attention networks, the head number was 4. In the position-wise feed-forward networks, the inner dimension was 2048. Besides, we trained a 2-layer 1024-dimensional LSTM network on HKUST transcriptions as the external language model and adopted the above 3655-sized vocabulary set. \par
During training, we used the CTC/attention joint training ($\alpha=0.7$) and the Adam optimizer with Noam learning rate schedule (25000 warm steps)\cite{Vaswani2017}, and trained for 30 epochs. To prevent overfitting, we used dropout \cite{Nitish2014dropout} (dropout rate $\!=\!0.1$) in each sub-layer, uniform label smoothing \cite{44903} (penalty $\!=\!0.1$) in the output layer and the model averaging approach that averages the parameters of models at the last 10 epochs. During decoding, we adopted online joint decoding approach, combining T-CTC prefix scores ($\lambda\!=\!0.5$) and language model scores ($\gamma\!=\!0.3$) to prune the hypotheses, and the beam size was 10. \par
\subsection{Chunk-SAE with or without reusing states}
\label{ssec:res-1}
In Table~\ref{tab:TE}, we compared the speed and performance of the chunk-SAE with or without reusing states. The context configuration remained the same for online models during the comparison, i.e. $N_l\!=\!N_c\!=\!N_r\!=\!64$. Firstly, we measured the speed of various encoders during decoding using a sever with Intel(R) Xeon(R) Silver 4114 CPU, 2.20GHz. For the clear comparison, we set the speed of chunk-SAE to $1.0$ and give the speed ratio of other encoders. In lines 1 and 2 of Table~\ref{tab:TE}, the chunk-SAE was slower than the SAE due to the redundant computation of the historical and future context. In lines 2 and 3 of Table~\ref{tab:TE}, we observed that the state reuse chunk-SAE was 1.5x faster than the chunk-SAE, which is consistent with the theoretical analysis in Section~\ref{ssec:c-sae}. In addition to the faster speed, the state reuse chunk-SAE outperformed the chunk-SAE by $1.53\%$ and $0.38\%$ relative CERs reduction on HKUST development and test set, respectively. Because of the faster speed and better performance, we employed the state reuse chunk-SAE in our subsequent experiments. \par

\begin{table}
  \caption{The character error rates (CER) of different Transformer-based ASR models on HKUST.}
  \label{tab:TE}
  \centering
  \begin{tabular}{c c c c c c}
    \toprule
    \multirow{2}{*}{Encoder} &
    \multirow{2}{*}{Decoder} &
    State &
	Encoder &
    \multirow{2}{*}{Dev} &
    \multirow{2}{*}{Test}\\
    & & Reuse & Speed Ratio& & \\
    \midrule
    SAE & SAD & -- & 2.8 & 24.12 & 23.47\\
    \midrule
    Chunk- & MTA- & $\times$ & 1.0 & 24.83 & 23.74\\
    \cmidrule{3-6}
    SAE & SAD & $\surd$ & 1.5 & 24.45 & 23.65\\
    \bottomrule
  \end{tabular}
\end{table}

\subsection{Context investigation}
\label{ssec:res-3}
in Table~\ref{tab:N}, we investigated our online model performance varying the historical, central and future context lengths. Firstly, comparing lines 2-4 in Table~\ref{tab:N}, we can see that the future context brought more improvement than the historical context, which indicates that the future context is more crucial to the performance of our online models. Secondly, comparing lines 5-7 in Table~\ref{tab:N}, we found that it was effective to increase the length of the historical context when we intended to reduce the latency of the state reuse chunk-SAE and maintain the recognition accuracy at the same time. Thirdly, comparing lines 7 and 8 in Table~\ref{tab:N}, we found that the CER reduced when we increased the length of central context. \par
\begin{table}
  \caption{The CERs of online Transformer-based ASR models with different context configurations on HKUST.}
  \label{tab:N}
  \centering
  \begin{tabular}{c c c c c c c}
    \toprule
    No. & Model & $N_l$ & $N_c$ & $N_r$ & Dev & Test \\
    \midrule
    1 & SAE+SAD & -- & -- & -- & 24.12 & 23.47\\
    \midrule
    2 & & 0 & 64 & 0 & 30.02 & 28.53\\
    3 & State Reuse & 64 & 64 & 0 & 29.97 & 28.41\\
    4 & Chunk-SAE & 0 & 64 & 64 & 24.94 & 24.10\\
    5 & +  & 64 & 64 & 64 & \textbf{24.45} & \textbf{23.65}\\
    6 & MTA-SAD & 64 & 64 & 32 & 24.67 & 24.05\\
    7 &  & 96 & 64 & 32 & 24.50 & 23.66\\
    8 & & 128 & 32 & 32 & 25.04 & 24.21\\
    \bottomrule
  \end{tabular}
\end{table}
\begin{table}
  \caption{Comparison with published ASR models on HKUST.}
  \label{tab:pub_online}
  \centering
  \begin{threeparttable} 
  \begin{tabular}{l c c}
    \toprule
    Model  & Size & Test\\
    \midrule
    TDNN-hybrid, lattice-free MMI \cite{Povey2016PurelySN}  & 19M & 23.69\\
	Offline Self-attention Aligner \cite{8682954}  & 38M\tnote{*} & 24.12\\
	Online Self-attention Aligner \cite{8682954}  & 24M\tnote{*} & 26.52\\
	Offline BLSTM CTC/att model \cite{watanabe2017hybrid}  & 112M & 27.43\\
    Online LC-BLSTM CTC/att model \cite{mta} & 112M & 27.84\\
    Online Transformer-based CTC/att model & 31M & \textbf{23.66}\\
    \bottomrule
  \end{tabular}
\begin{tablenotes}  
\item[*] \footnotesize Estimated model parameter size according to model configurations.
\end{tablenotes}  
\end{threeparttable}  
\end{table}
Finally, our best online model achieved a $23.65\%$ CER, with a 640 ms latency and a $0.18\%$ absolute CER degradation compared with the offline baseline in line 1 of Table~\ref{tab:N}. In Table~\ref{tab:pub_online}, we also compared our Transformer-based online CTC/attention model with other published ASR models. For a fair comparison, the latency of the online E2E models listed in Table~\ref{tab:pub_online} is 320 ms. These models were trained on HKUST with speed perturb except online Self-attention Aligner model. \par

\section{Conclusion}
\label{sec:con}

In this paper, we propose the Transformer-based online E2E ASR model, which consists of the state reuse chunk-SAE and MTA based SAD, and integrate the proposed Transformer-based online E2E ASR model into the CTC/attention ASR architecture. Compared with the simple chunk-SAE, the state reuse chunk-SAE performs better and requires less computational cost, because it has broader historical context via storing the states in previous chunks.
Compared with the SAD, the MTA based SAD truncates the SAE outputs in a monotonic left-to-right way and performs attention on the truncated SAE outputs, making it applicable to online recognition.
We evaluate the proposed Transformer-based online CTC/attention E2E models on HKUST and achieves 
a $23.66\%$ CER with a 320 ms latency, which outperforms our prior LSTM-based online E2E models. In future, we plan to adopt teacher-student learning approach to further reduce the model latency.\par
%

\vfill
\pagebreak

\bibliographystyle{IEEEbib}
\bibliography{strings,refs}

\end{document}